\documentclass{vgtc}                 

\ifpdf
  \pdfoutput=1\relax                   
  \usepackage{graphicx}                
  \DeclareGraphicsExtensions{.pdf,.png,.jpg,.jpeg} 
\else
  \usepackage{graphicx}                
  \DeclareGraphicsExtensions{.eps}     
\fi%

\graphicspath{{figures/}{pictures/}{images/}{./}} 

\usepackage{microtype}                 
\PassOptionsToPackage{warn}{textcomp}  
\usepackage{textcomp}                  
\usepackage{mathptmx}                  
\usepackage{times}                     
\usepackage{cite}                      
\usepackage{tabu}                      
\usepackage{booktabs}                  

\usepackage{amsmath,bm}

\usepackage{array,multirow,graphicx}
\usepackage[table,xcdraw]{xcolor}

\usepackage{xcolor}
\usepackage{color}
\usepackage{enumitem}
\usepackage{textcomp}
\definecolor{orange}{rgb}{1,0.5,0}

\onlineid{3735}

\vgtccategory{Research}

\vgtcinsertpkg

\title{Investigating Psychological Ownership in a Shared AR Space: \\Effects of Human and Object Reality and Object Controllability}

\author{Dongyun Han\thanks{e-mail: dongyun.han@usu.edu}\\ %
        \scriptsize Utah State University %
\and Donghoon Kim\thanks{e-mail: donghoon.kim@usu.edu}\\ %
     \scriptsize Utah State University %
\and Kangsoo Kim\thanks{e-mail: kangsoo.kim@ucalgary.ca}\\ %
     \scriptsize University of Calgary %
\and Isaac Cho\thanks{e-mail: isaac.cho@usu.edu}\\ %
     {\scriptsize Utah State University }}

\teaser{
  \centering
    \includegraphics[width=.85\textwidth]
    {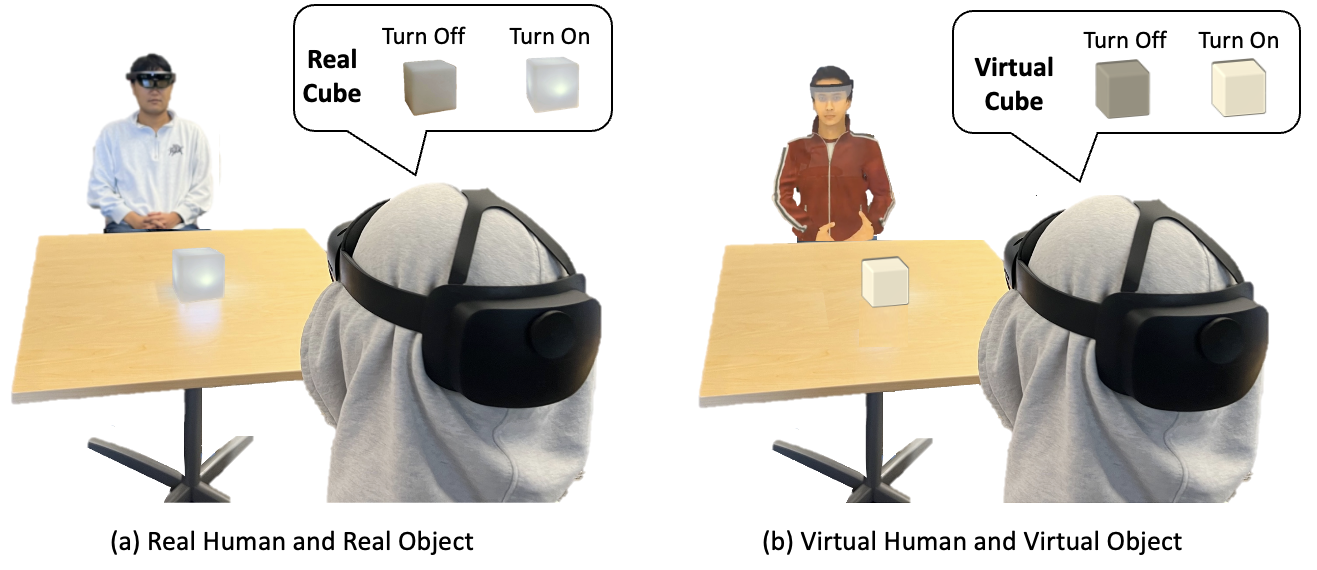}
    \caption{Study environment. Participants interact with a real or virtual cube object by using voice commands to turn on and off the light with either a real or virtual partner. (a) shows a scenario in which a participant and a real human partner interact with a real cube object in a shared space. (b) depicts a scenario in which a participant and a virtual partner interact with a virtual object in a shared AR setting. The interaction scenarios are determined by the partner and object reality, and object controllability.
    } 
    \label{fig_teaser}
}

\abstract{
Augmented reality (AR) provides users with a unique social space where virtual objects are natural parts of the real world. 
The users can interact with 3D virtual objects and virtual humans projected onto the physical environment. 
This work examines perceived ownership based on the \emph{reality} of objects and partners, as well as object \emph{controllability} in a shared AR setting. 
Our formal user study with 28 participants shows a sense of possession, control, separation, and partner presence affect their perceived ownership of a shared object.
Finally, we discuss the findings and present a conclusion.
} 

\CCScatlist{
  \CCScatTwelve{Human-centered computing}{Visu\-al\-iza\-tion}{Visu\-al\-iza\-tion techniques}{Treemaps};
  \CCScatTwelve{Human-centered computing}{Visu\-al\-iza\-tion}{Visualization design and evaluation methods}{}
}

\begin{document}


\firstsection{Introduction}

\maketitle

With the advancement of augmented reality (AR) devices and technologies, AR environments have the potential to become spaces where multiple users engage in social interaction, such as avatar-mediated remote collaboration and communication~\cite{billinghurst1999collaborative, 
he2020collabovr}.
In such social settings, users often share and interact with virtual or real objects to perform a given task for common goals collaboratively or compete with each other.
Here, understanding the perceived ownership of an object, i.e., who owns or controls what, is important for establishing a common ground, increasing the quality of communication, and achieving the interaction goals.
Previous research studied the ownership of virtual objects in immersive virtual environments~\cite{poretski2019virtual, poretski2021owns,carrozzi2019s}, but it still lacks a comprehensive understanding of virtual/real object ownership associated with the users' interaction settings, especially in a shared AR space.

This paper tackles the research gap by conducting a human-subjects study that investigates the effects of different \emph{reality} and \emph{controllability} settings on the perceived ownership of a virtual or real object in a shared AR environment.
In the study, participants are provided with an object (a controllable light cube), which is shared with an experimental confederate who plays a role of an interaction partner.
The \emph{reality} conditions of both object and confederate vary between virtual and real forms---i.e., a virtual vs. real object, and a virtual vs. real human.
For the \emph{controllability} conditions, participants and the confederate may or may not have the authority to turn on/off the object.
While varying these \emph{reality} and \emph{controllability} settings in our study conditions, we measure the participants' subjective perception of ownership of the object and the sense of presence with the object and confederate.
Our results show that the object-reality and the user-controllability influence the overall ownership perception significantly.
Finally, We discuss the implications of these findings in social interaction contexts while justifying them with previous findings and knowledge in psychology and perception science as well as the participants' feedback after the experiment.

\section{Related Work}
\label{sec_relatedWork}

\subsection{Psychological Ownership of Virtual Objects in AR}

The traditional definition of psychological ownership describes it as a state of mind in which an individual feels that an object is his or her own~\cite{pierce2001toward}. 
Among various factors that could contribute to the perceived ownership of objects, five major factors were identified in the literature: (1) possession, (2) control, (3) identity, (4) responsibility, and (5) territoriality.
Possession refers to the sense that an item or entity is one's own~\cite{pierce2003state}.
Control is the extent to which a person can use personal initiative in tasks linked with the target object~\cite{avey2009psychological, olckers2013psychological}. 
Identity is the cognitive link between an individual and the target object, e.g., police officer and gun~\cite{avey2009psychological, buchem2012psychological}.
Responsibility is the sense of explicit liabilities associated with that target object~\cite{avey2009psychological, olckers2013psychological, van2004psychological}. 
Territoriality is a physiological space like a personal space where one conducts protective behavior towards the target object from others, in other words, a sense of separation~\cite{avey2009psychological, van2004psychological, rudmin1987semantics}.

Previous research and applications have shown that AR users could feel a sense of ownership over virtual objects or creatures even though they are not real.
For example, Poretski et al.~\cite{poretski2019virtual} evaluated an experience of interacting with a virtual dog on a mobile AR application for three weeks, and found that they felt closeness and responsibility for the virtual dog.
Pokemon Go~\cite{PokemonGo} is an exemplary application that encourages users to compete with each other and build ownership of virtual creatures by collecting or breeding stronger AR creatures.
In a shared AR setting where multiple users are co-located with a shared virtual object, Poretski et al.~\cite{poretski2021owns} investigated the relative ownership of virtual objects between two participants in different interaction scenarios, such as creating, manipulating, and editing virtual objects.
They found that the user who created the object and had it within their personal space tended to argue the ownership of the object.
Carrozzi et al.~\cite{carrozzi2019s} demonstrated that the controllability of an AR object's color and position was associated with the perceived ownership and affected the users' consumption experience in an online marketing context.

However, it is unclear how the perceived ownership could be affected by the reality state of objects (i.e., real objects vs. virtual objects). 
It is also not explored well how the presence of other users and their controllability towards the shared object can influence the perceived ownership in a shared AR.

\subsection{Virtual Humans and Social Presence in AR}

Virtual humans refer to a human-like computer graphic representation, e.g., embodied virtual avatars and agents, which are typically used in AR-mediated multi-user interactions. 
Due to their human-like appearance and interactive behaviors, people naturally distinguish them from non-human 3D objects and often treat them as if they are real humans with a sense of social/co-presence~\cite{sanz2015virtual, llobera2010proxemics}--a sense of being together with other persons.
Previous research has shown that a high sense of social presence is positively associated with various perception and performance measures~\cite{piumsomboon2018mini}, and possibly influences the sense of object ownership~\cite{VenusJin2023spa}.

Researchers have tried to increase the level of social presence with virtual humans.
For example, Bailenson et al.~\cite{bailenson2001equilibrium} showed that virtual humans with natural behaviors like gaze and body movements had a greater sense of presence with participants. 
The sense of presence with virtual humans could also increase when they interacted with actual objects in the physical world.
Similarly, Kim et al.~\cite{kim2017exploring} investigated the effects of visual conflicts where a virtual human or object shared the same space with a physical object, during social interactions, and found that a higher social presence with the virtual human was reported when she avoided visual conflict with the surrounding physical objects, such as tables or chairs. 
Lee et al.~\cite{lee2019mixed, lee2016wobbly} demonstrated that participants experience a greater sense of presence when a virtual human could physically interact with the surrounding environment by employing a controllable wobbly table and movable objects on a table.
Based on the findings of previous studies, in this paper, we investigate the potential associations of the real/virtual human interaction partner on the ownership of a shared real/virtual object.

\section{Experiment}


Our research aims to investigate the psychological ownership of a virtual or real object when it is shared with a virtual or real human in a shared AR setting, while broadly covering the following research questions (RQs):

\begin{enumerate}
[label=\textbf{RQ\arabic*.}, leftmargin=0.35in]
\setlength\itemsep{0pt}
\item How does the \emph{reality} of an object and interaction partner (i.e., real vs. virtual) affect a user's psychological ownership of the object?



\item How does the ability to control an object (\emph{controllability}) affect a user's perceived ownership? 

\item Are there any interaction effects among these \emph{reality} and \emph{controllability} factors?
\end{enumerate}

\noindent To address these questions, we conducted a human-subject experiment, which is detailed in this section.

\subsection{Study Design and Hypothesis}
\label{Sec:StudyDesign}

For the study, we used a within-subjects design based on four factors related to \emph{reality} and \emph{controllability}:



\begin{enumerate}[label=-,leftmargin=0.1in]
\setlength\itemsep{0pt}
    \item \textbf{Object Reality (OR)}: The controllable object is either real or virtual cube (\textit{RCube} vs. \textit{VCube}).
    \item \textbf{Partner Reality (PR)}: The interaction partner is either real or virtual human (\textit{RHuman} vs. \textit{VHuman}).
    \item \textbf{User Controllability (UC)}: The participant can or cannot control the object (\textit{UCont} vs. \textit{UNoCont}).
    \item \textbf{Partner Controllability (PC)}: The interaction partner can or cannot control the object (\textit{PCont} vs. \textit{PNoCont}).
\end{enumerate}





\noindent Given these factors, we established three hypotheses as follows:

\begin{enumerate}
[label=\textbf{H\arabic*.}]
\setlength\itemsep{0pt}
    \item The participant will have a stronger sense of ownership factors when the tangibility of a shared object is real rather than virtual. 
    \item The participant's ownership factors of the shared object will be hampered when the tangibility of the partner is a real human rather than a virtual human. 
    \item Having control over an object will reinforce the perceived ownership factors of the object.
\end{enumerate}

\subsection{Participants}
We recruited 28 participants (15 male, 13 female; average age of 20.0 years ranging from 18 to 24) via SONA. 
All participants had (normal/corrected) 20/20 vision.
The study took approximately 30 minutes, and the participants were given 1.5 SONA credits as a reward, which was the maximum reward for a 30-minute in-person study in accordance with IRB regulations  (IRB \#13233).

\subsection{Interaction Session and Apparatus}
\label{section_task}


In our study, two researchers played two different roles: one as an experimenter who guided participants and provided instruction, and the other as an interaction partner.
The interaction partner shared the experiment space with the participants and performed the object-controlling tasks together.
The task involved two people: a participant and the interaction partner---the partner could be a real or virtual human depending on the study conditions. 
We chose a virtual character resembling the study confederate as seen in Figure~\ref{fig_teaser} from the Microsoft Rocketbox Avatars library~\cite{gonzalez2020rocketbox} to serve as a virtual partner in our study. This is to minimize the impacts of different avatar appearances on perceived ownership. They were seated across the table, on which a real/virtual light cube object was located in the center of the table as shown in Figure~\ref{fig_teaser}. The real light cube object measures 4 inches in height, width, and depth, matching the dimensions of the virtual object.

\begin{figure}[t]
    \centering
    \includegraphics[width=\linewidth]{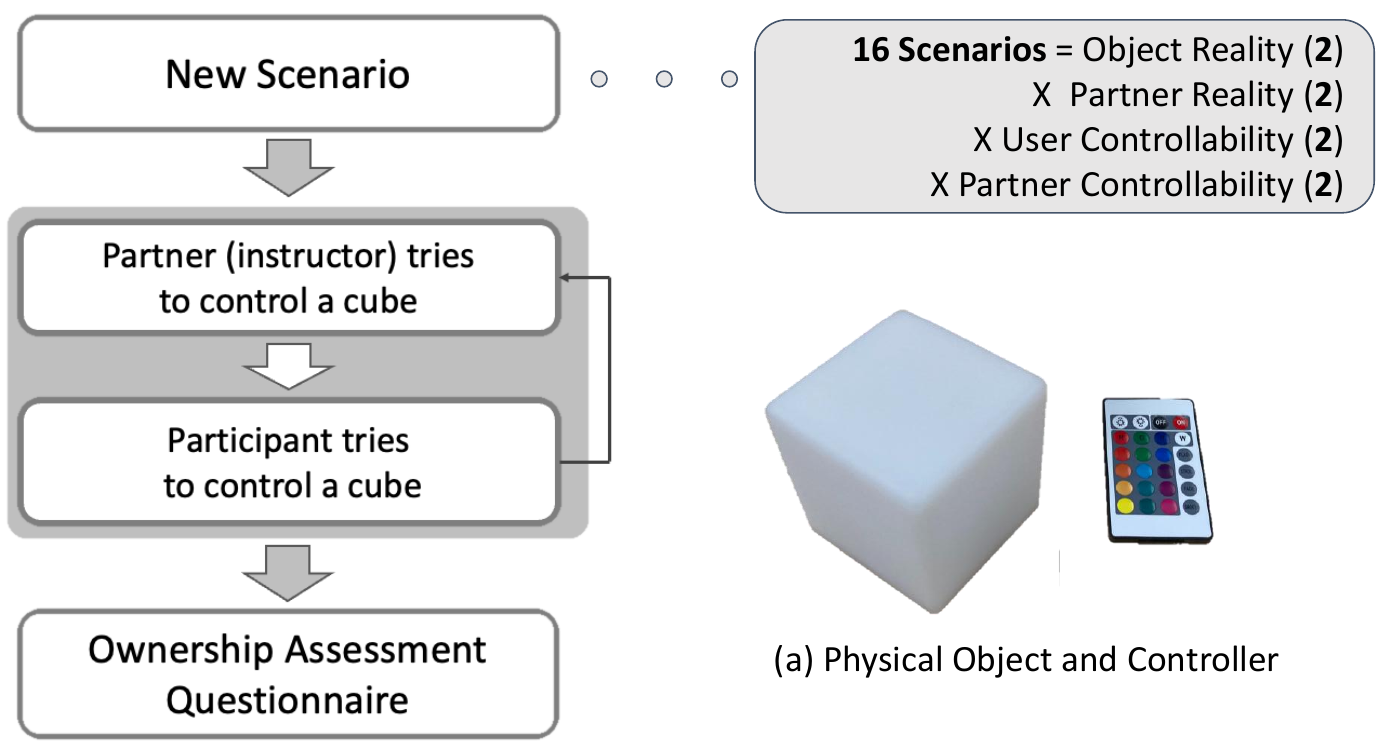}
    \caption{The study procedure. There were a total of 16 different study settings based on the four factors of reality and controllability.
    }
    \label{Fig:procedures}
    \vspace{-.5cm}
\end{figure}

During the study, participants wore an optical see-through AR headset (Microsoft HoloLens 2), which provides a 50-degree field of view, a resolution of 2048$\times$1080 for each eye, and a refresh rate of 75\,Hz.
The interaction partner wore another headset (Microsoft HoloLens 1) pretending to see the same scene as the participants. 
We used a Wizard of Oz paradigm~\cite{kelley1983empirical} where the experimenter actually controlled the real/virtual object using a remote controller (Figure~\ref{Fig:procedures}a) and a wireless keyboard, while the participant and partner performed the task.

In the study, the participant and the partner tried to control the light of the cube by voice commands speaking ``Turn On'' or ``Turn Off''.
The tasks consisted of a total of 16 interaction sessions due to the number of study condition combinations (2 OR $\times$ 2 PR $\times$ 2 UC $\times$ 2 PC).
For example, when the condition involved a virtual object and/or a virtual human confederate, the AR headset visualized them at the same locations where the real counterparts were located (Figure~\ref{fig_teaser}b). During the interaction, the virtual partner's gaze altered between the cube object and the participant's head. The gaze target was changed every 3--5 seconds to give the participant the impression that the virtual partner recognized the participant. Similarly, when the real partner engaged in the interaction, he mimicked these actions. 
When the virtual partner spoke the voice commands, his lips moved in response to the pronunciation of the commands using the LipSync plugin~\cite{libSyncPro}. 
The virtual scene was developed using Unity 2020.3.43f1.

\subsection{Procedure}

Once a participant arrived and signed an informed consent form, we explained our study objectives and procedure.
The experimenter guided the participant and the study confederate (i.e., interaction partner) to sit on chairs across a table shown in Figure~\ref{fig_teaser}. Before beginning the main study, the participant had a training session with the real/virtual partners. He or she learned how to control the real/virtual object with their voice. Following the training session, the participant proceeded to the main study.

The main study includes 16 interaction sessions (2 OR $\times$ 2 PR $\times$ 2 UC $\times$ 2 PC). 
Only one combination of the real/virtual partners and cubes was part of a session, and the other partner and cube were moved out of the participant's sight. For each session, the participant and a partner performed the object manipulation interaction four times (i.e., turning the light cube on and off twice). 
The controlling orders were randomly determined and instructed by the experimenter. It took about 30 seconds. 
After the interaction, the participant was asked to answer a questionnaire regarding the perceived ownership and the sense of presence with the object, taking 60 seconds. 
After completing all the sessions, the participant was guided to complete a post-questionnaire as the last step of this study, asking their overall perceptions comparing all the sessions that they experienced. 
The experiment was terminated once the participant completed the final questionnaire.


\subsection{Measures}

For the study, we measured ownership factors to understand the effects of object/partner reality and user/partner controllability.
Also, we collected the participants' presence perception with the object and the partner to investigate the potential relationship with the perceived ownership.
Table~\ref{table:questions} shows the seven questions used in the study, which are also described below. 
The participants were asked to rate each factor on a 7-point Likert scale (1: not at all -- 7: very much) after experiencing each interaction session. 

\paragraph{Ownership Factors:} 
Among the five factors contributing to the perceived ownership described in Section~\ref{sec_relatedWork}, we decided to use four factors---\textit{Possession}, \textit{Control}, \textit{Identity}, and Territoriality---excluding Responsibility due to the irrelevance to our study setting. 
In this work, we use the term, \textit{Separation}, instead of territoriality emphasizing the perceived distinction of the real and virtual worlds (\textbf{Q1--Q4} in Table~\ref{table:questions}). 

\begin{table}[t] 
\caption{\label{table:questions} Measures and the questions used in the study.}
\begin{tabular} {m{0.6cm} | m{1.8cm} | m{5cm}}
\toprule 
    {Q1} & {Possession}  & {To what extent do you feel a sense of belongingness to the object in this session?}\\
\toprule 
    {Q2} & {Control}  & {To what extent do you feel the right to control the object in this session?}\\
\toprule 
    {Q3} & {Identity}  & {To what extent do you feel a bond to the object in this session?}\\
\toprule 
    {Q4} & {Separation}  & {To what extent do you feel separate/distinct from the object in this session?}\\
\toprule 
    {Q5/6} & {Object/Partner Presence}  & {During
the session, to what extent did you have a sense of being in the same room with your object/partner?}\\
\toprule 
    {Q7} & {Ownership}  & {Who do you think own the object?}\\
\toprule 
\end{tabular}
\end{table}

\paragraph{Object/Partner Presence:} To investigate the possible relationship between the (social) presence of the object/partner and the perceived ownership, we included two presence-related questions on the same 7-point Likert scale (\textbf{Q5/Q6} in Table~\ref{table:questions}). 

\paragraph{Perceived Ownership:} In addition to the ownership factors and presences, we added a binary choice question asking whether the virtual or real object was perceived to belong to either the participant or the partner (\textbf{Q7} in Table~\ref{table:questions}). To determine which factors strongly influence the participants' perceived ownership over the objects, we avoid providing ambiguous answers that both and neither owned the object. 




\paragraph{Post-Study Feedback}
After all sessions, we collected qualitative feedback about how the object/partner reality and object controllability affected the perception of ownership using open questions in the post-questionnaire as follows: 

\begin{enumerate}[label=-,leftmargin=0.1in]
\setlength\itemsep{0pt}
    \item How does the object reality (RCube cube vs. VCube) affect your perceived ownership of the object?
    \item How does the human reality (RHuman vs. VHuman) affect your perceived ownership of the object?
    \item How does object controllability affect your perceived ownership of the object?
   
\end{enumerate}

\begin{figure*}[t!]
    \centering
    \includegraphics[width=\linewidth]{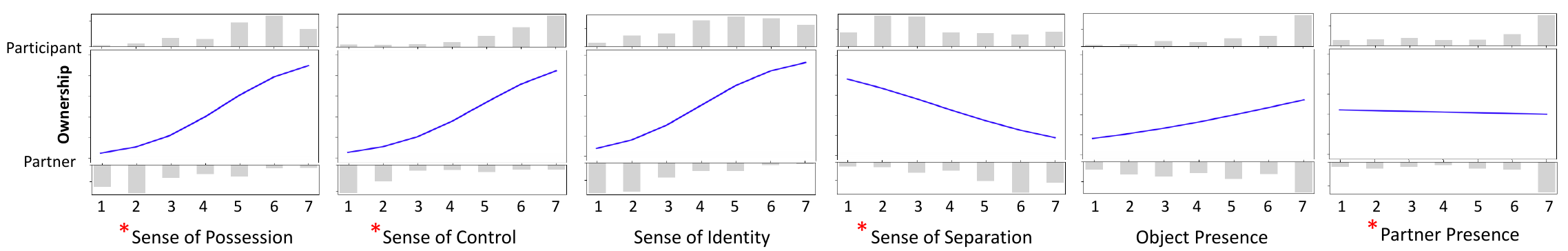}
    \caption{Relationships between the perceived ownership (i.e., participants vs. partner) and ownership factors or presence measurements. (*) indicates the measurements having the statistical effects in Section~\ref{sec_logistic_results}. 
    The histograms placed above and below show the regression results that depict the relative distribution of responses by ownership and measurement scores. 
    }
    \label{fig_regression_result}
    \vspace{-3ex}
\end{figure*}

\section{Results}

Statistical analyses including multivariate logistic regression analysis and 4-way repeated ANOVA were performed to analyze the measurements. 
We used scikit-learn~\cite{scikit-learn}, a Python package, to fit a logistics regression model.
This extracted what measurements were highly relevant to the perceived ownership.
In this section, multivariate logistic regression results are first reported to examine the effect of the ownership factors and presence on the user-perceived ownership of the objects.
Next, we report the ANOVA results regarding the highly relevant measurements we found from the logistic regression results.
The four independent variables are OR, PR, UC, and PC.
The Bonferroni pairwise comparison test was used as a post-hoc test when ANOVA showed a significant effect. 
The significance level was set at 5\%.

\subsection{Regression Results on Ownership}
\label{sec_logistic_results}

We found a statistically significant relationship of the perceived ownership with the sense of possession ($p=.002$), control ($p<.001$), and separation $(p<.001$), and partner reality ($p<.001$), but not with the sense of identity ($p=.095$) and object reality ($p=.988$).
The results are reported in Figure~\ref{fig_regression_result}.

To quantify the strength of the association between possession, control, separation, and partner reality to ownership, we report the odds ratio for each variable.
It is defined as the likelihood of an outcome occurring given a specific exposure, as opposed to the likelihood of the outcome occurring in the absence of that exposure. 
The odds of ownership increased by 51\% (95\% CI [.17, .96]) and 68\% (95\% CI [.38,.1.05]) for the possession and control and decreased by 35\% (95\% CI [-.42, -.27]) and 26\% (95\% CI [-.34, -.15]) for the separation and partner reality, respectively.

\subsection{ANOVA Results on Ownership Factors and Presence}
\label{sec_anova_results}
We further conducted ANOVAs for the four measurements that showed significant relationships in the regression analysis.
Table~\ref{result_table} shows all the significant main and interaction effects of the four measurements found in Section~\ref{sec_logistic_results}, and the details are described below.

\paragraph{Sense of Possession:} 
The UC and PC factors showed a significant interaction effect ($p<.001$). 
The simple impacts of UC were identified in both PCont and PNoCont, reporting ($p<.001$, $F(1, 27)=102$, $\eta^2_p=0.790$) and ($p<.001$, $F(1, 27)=118$, $\eta^2_p=0.814$) respectively.
The former shows that the participants felt higher possession in UCont ($M=4.39$) than in UNoCont ($M=2.11$). 
In the latter, UCont ($M=5.07$) also has a higher sense of possession than UNoCont ($M=2.04$). 


The main effects of OR ($p<.001$), UC ($p<.001$), and PC ($p=.003$) were also found. 
The sense of possession was significantly higher when dealing with RCube ($M=4.00$) than VCube ($M=3.60$).
Participants also felt a higher sense of possession when they had controllability (UCont, $M=5.11$) than UNoCont ($M=2.50$). 
When the partner had controllability (PCont, $M=3.59$), the perceived belonging was lower than PNoCont ($M=4.02$).
These results support \textbf{H1} and \textbf{H3}, but not H2.

\paragraph{Sense of Control:}
We found an interaction effect between UC and PC ($p=.011$).
The simple effects of UC were discovered in both cases of PCont ($p<.001$, $F(1, 27)=133$, ${\eta_p}^2=0.831$)  and PNoCont($p<.001$, $F(1, 27)=305$,  ${\eta_p}^2=0.919$). 
In the PCont conditions, the participants showed a greater sense of control in UCont ($M=5.45$) than UNoCont ($M=2.05$).
The PNoCont conditions also showed similar results. 
A greater sense of control was found when the participants had control (UCont, $M=6.15$) than UNoCont ($M=2.10$). 
These results support \textbf{H3}. 
In our further analysis, the sense of control was significantly higher when only the participants have control ($M=6.15$) than both have it ($M=5.45$).

We also found the main effects of OR ($p=.041$), UC ($p<.001$), and PC ($p=.008$).
The sense of control over RCube ($M=4.04$) was statistically greater than the sense over VCube ($M=3.82$), supporting \textbf{H1}.
UCont ($M=5.80$) had a greater sense of control than UNoCont ($M=2.06$). 
When the partner had controllability (PCont, $M=3.73$), a weaker sense of control was reported than PNoCont ($M=4.13$). 
No evidence for H2 was found.



\paragraph{Sense of Separation:}
We found an interaction effect between OR and PC ($p<.001$). 
Only a simple effect of PC with the VCube condition was disclosed ($p<.001$, $F(1, 27)=18.1$, ${\eta_p}^2=0.402$). 
When the object was virtual, PCont ($M=4.88$) raised the sense of separation from the object more than PNoCont ($M=4.34$). 

Significant main effects of OR ($p=.036$) and UC ($p<.001$) were reported.
VCube ($M=4.62$) strengthened the sense of separation more than RCube ($M=4.35$). 
This result supports our \textbf{H1}.
In addition, the sense of separation was intensified with UNoCont ($M=5.32$) compared to with UCont ($M=3.64$), supporting \textbf{H3}. 
We did not find any evidence to support H2.

\bgroup
\def\arraystretch{1.2}%
\begin{table}[t]
\centering
\normalsize
\caption{\label{result_table} Interaction and main effects on the Sense of Possession, Control, Separation, and Presence.}
\begin{tabular} {m{2cm} | m{1.3cm} | m{1cm} | m{0.9cm}| m{0.7cm}}
\toprule
\textbf{Measures} & \textbf{Factor}  &  {$\bm{F(1, 27)}$} & {$\bm{p}$} & {$\bm{\eta_p^2}$}\\
\midrule
   & {UC \texttimes{} PC}     &  {14.0} & {$<.001$}  &  {.342}  \\ \cline{2-5}
  \textbf{Sense of}   & {OR}   &   {15.7}  & {$<.001$}  & {.369}    \\ \cline{2-5}
  \textbf{Possession}  &  {UC}      &  {108}  & {$<.001$}  & {.800}   \\ \cline{2-5}
   &  {PC}      &  {10.5}  & {.003}  & {.280}   \\
\midrule
                    & {UC \texttimes{} PC}      &  {7.38}  &  {.011} &  {.215}  \\ \cline{2-5}
\textbf{Sense of}   & {OR}      &  {4.61} &  {.041} &   {.146} \\ \cline{2-5}
\textbf{Control}    & {UC}      & 244  & {$<.001$}  &  .900 \\ \cline{2-5}
                    & {PC}    &  8.14  & .008  &  .232  \\ 
\midrule
\textbf{Sense of}   & {OR \texttimes{} PC}      &  {13.7}  &  {$<.001$} &  {.337}  \\ \cline{2-5}
\textbf{Separation} & {OR}      &  {4.88} &  {.036} &   {.153} \\ \cline{2-5}
                    & {UC}      &  {32.1}  & {$<.001$}  &  {.543}  \\ 
\midrule

\textbf{Partner }   & {UC \texttimes{} PC}      &  {4.34}  &  {.047} &  {.139}  \\ \cline{2-5}
\textbf{Presence}  & {PR}      &  {84.6} &  {$<.001$} &   {.758} \\ \cline{2-5}
  & {PC}      &  {6.59}  & {.016}  &  {.196}  \\ 
\bottomrule
\end{tabular}
\vspace{-2ex}
\end{table}



\paragraph{Partner Presence:} 
The UC and PC factors had a significant interaction effect ($p=.047$).
When the participants had controllability, a simple impact of PC was found ($p=.002$, $F(1, 27)=11.7$, $\eta_p^2=0.303$).
Interestingly, when the partner could control the object (PCont, $M=5.52$), the perceived partner presence was improved much higher than PNoCont ($M=5.14$).

We found main effects of PR ($p<.001$) and PC ($p=.016$).
RHuman ($M=6.68$) had a greater partner presence than VHuman ($M=6.68$), supporting \textbf{H2}.
Also, participants perceived a better partner presence in UCont ($M=5.34$) condition than in UNoCont ($M=5.10$), supporting \textbf{H3}. 
We found no evidence for H1.


\section{Discussion}

\paragraph{\textbf{Having control over objects has a significant impact on the sense of ownership.}}
The interaction effects on the sense of possession and control indicate that the participants who have control over an object have a stronger sense of ownership, regardless of whether their partner has control or not. 
Additionally, the results about the sense of separation show that the participants experienced less separation when they had control over it. 
These findings are consistent with the participants' feedback. 
Regardless of the object's realism, the majority of participants said they felt a stronger sense of ownership when they were the only ones who had control over it. 
However, a few participants expressed difficulty determining who owned it when both themselves and their partners could control or could not control the object.

\paragraph{\textbf{Real objects can induce a greater sense of ownership than when they are virtual.}}

Our results indicate that when objects are real, users can have a stronger sense of possession and control and feel less separated from the objects, regardless of whether the interaction partner has control over them or not.
When the cube was virtual in our study, however, it was confirmed that the separation was felt more when the partner was in control over the object.
A few participants responded that they felt they owned the real object more than the virtual object due to the object's realism.
They stated that the real object was more likely to belong to a real human or real partner, whereas the virtual cube belonged to a virtual person.

\paragraph{\textbf{Mutual controllability over a shared object can increase the sense of the (social) presence of the interaction partner.}}

Not surprisingly, it was confirmed that the presence of the partner increased when the partner was a real human. 
We found no difference in the partner presence whether the partner had control or not when the participants did not have control over the object.
However, interestingly, when both the participants and the partner had control over the object, there was a significantly higher level of partner presence reported.
This finding is explained by Carrozzi et al.~\cite{carrozzi2019s}, revealing that distinct interactions with a virtual object allow users to maintain social differentiation while also assimilating with peers.

\paragraph{\textbf{Partner reality (or a strong sense of partner presence) could possibly impact the perceived ownership.}}
In the ANOVA results, we did not find any strong effects of the partner reality on the ownership factors, but there was a significant effect on the perceived presence of the partner. 
According to the logistic regression results, the partner presence had a negative correlation with the perceived ownership---in other words, a strong sense of partner presence could reduce the user's ownership of the shared object.
These findings imply that the partner reality has a negative impact on ownership.
This result, however, should be interpreted with caution because it may be related to the limitations of our study, which we discuss more in the following section.



\subsection{Limitations and Future Work}
Some limitations could stem inherently from the nature of the study design and the way to measure and collect the subjective perception.
The study highlighted the challenges that participants faced in determining ownership when both themselves and the real human partner had control over the object.
Some participants commented that they thought the real human partner most likely owned the object because they perceived he was a part of the study.
Further experiments are required to thoroughly investigate the effect of a partner's presence, particularly if the partner is a real human.

In the study, the object-controlling tasks were bland and the duration of interaction was relatively short.
One participant stated that as the study progressed, he began to feel a sense of possession of the virtual object.
This result might support Chalmers' argument~\cite{chalmers2017virtual} that virtual objects can be perceived as similar to real objects. However, a long-term effect of the interaction on perceived ownership, and its effect on engagements in a complicated collaborative task remains unexplored as our study did not intend to examine such a long-term effect. 
It is also unclear how the visual representation quality of a virtual object affects perceived ownership over time~\cite{chalmers2020reconstructing}. Future research could address the effect of long-term interaction on ownership, aiming to gain a deeper understanding of the topic.  


\section{Conclusion}

In this paper, we conducted a formal user study to examine how the perceived ownership differs by the object and interaction partner's reality states (i.e., real or virtual), and the object controllability in a shared AR setting.
Our results confirmed that three ownership factors (sense of possession, control, and separation) and partner presence are effective variables associated with the participants' perceived ownership of real and virtual objects. 
In our further analysis, we found that object reality and controllability have positive effects on the ownership factors, whereas partner reality does not.
Interestingly, a higher level of partner presence was reported when both the participants and the partner could control the object.
Finally, we discuss the potential limitations of our study as well as future research directions based on our findings and participant feedback.

\acknowledgments{
We acknowledge that this research is partially supported by the Natural Sciences and Engineering Research Council of Canada (NSERC), [RGPIN-2022-03294].
}

\bibliographystyle{abbrv-doi}

\bibliography{0_main_conf}

\begin{thebibliography}{10}

\bibitem{avey2009psychological}
J.~B. Avey, B.~J. Avolio, C.~D. Crossley, and F.~Luthans.
\newblock Psychological ownership: Theoretical extensions, measurement and
  relation to work outcomes.
\newblock {\em Journal of Organizational Behavior}, 30(2):173--191, 2009.

\bibitem{bailenson2001equilibrium}
J.~N. Bailenson, J.~Blascovich, A.~C. Beall, and J.~M. Loomis.
\newblock Equilibrium theory revisited: Mutual gaze and personal space in
  virtual environments.
\newblock {\em Presence: Teleoperators \& Virtual Environments},
  10(6):583--598, 2001.

\bibitem{billinghurst1999collaborative}
M.~Billinghurst and H.~Kato.
\newblock Collaborative mixed reality.
\newblock In {\em Proceedings of the International Symposium on Mixed Reality},
  pp. 261--284, 1999.

\bibitem{buchem2012psychological}
I.~Buchem.
\newblock Psychological ownership and personal learning environments: Do sense
  of ownership and control really matter.
\newblock In {\em PLE Conference Proceedings}, vol.~1. Citeseer, 2012.

\bibitem{carrozzi2019s}
A.~Carrozzi, M.~Chylinski, J.~Heller, T.~Hilken, D.~I. Keeling, and
  K.~de~Ruyter.
\newblock What's mine is a hologram? how shared augmented reality augments
  psychological ownership.
\newblock {\em Journal of Interactive Marketing}, 48(1):71--88, 2019.

\bibitem{chalmers2020reconstructing}
A.~Chalmers, J.~Zhao, D.~Medeiros, and T.~Rhee.
\newblock Reconstructing reflection maps using a stacked-cnn for mixed reality
  rendering.
\newblock {\em IEEE Transactions on Visualization and Computer Graphics},
  27(10):4073--4084, 2020.

\bibitem{chalmers2017virtual}
D.~J. Chalmers.
\newblock The virtual and the real.
\newblock {\em Disputatio: International Journal of Philosophy}, 9(46), 2017.

\bibitem{gonzalez2020rocketbox}
M.~Gonzalez-Franco et~al.
\newblock The rocketbox library and the utility of freely available rigged
  avatars.
\newblock {\em Frontiers in virtual reality}, 1:561558, 2020.

\bibitem{he2020collabovr}
Z.~He, R.~Du, and K.~Perlin.
\newblock {CollaboVR: A Reconfigurable Framework for Creative Collaboration in
  Virtual Reality}.
\newblock In {\em Proceedings of the IEEE International Symposium on Mixed and
  Augmented Reality}, pp. 542--554, 2020.

\bibitem{VenusJin2023spa}
S.~V. Jin and S.~Youn.
\newblock Social presence and imagery processing as predictors of chatbot
  continuance intention in human-ai-interaction.
\newblock {\em International Journal of Human–Computer Interaction},
  39(9):1874--1886, 2023. doi: {{%
10\hspace{.1pt}\discretionary{.}{%
}{.}\hspace{.4pt}1080\discretionary{/}{%
}{/}10447318\hspace{.1pt}\discretionary{.}{%
}{.}\hspace{.4pt}2022\hspace{.1pt}\discretionary{.}{%
}{.}\hspace{.4pt}2129277}}


\bibitem{kelley1983empirical}
J.~F. Kelley.
\newblock An empirical methodology for writing user-friendly natural language
  computer applications.
\newblock In {\em Proceedings of the CHI Conference on Human Factors in
  Computing Systems}, pp. 193--196, 1983.

\bibitem{kim2017exploring}
K.~Kim, G.~Bruder, and G.~Welch.
\newblock Exploring the effects of observed physicality conflicts on
  real-virtual human interaction in augmented reality.
\newblock In {\em Proceedings of the 23rd ACM Symposium on Virtual Reality
  Software and Technology}, pp. 1--7, 2017.

\bibitem{lee2016wobbly}
M.~Lee, K.~Kim, S.~Daher, A.~Raij, R.~Schubert, J.~Bailenson, and G.~Welch.
\newblock The wobbly table: Increased social presence via subtle incidental
  movement of a real-virtual table.
\newblock In {\em Proceedings of the IEEE Virtual Reality}, pp. 11--17, 2016.

\bibitem{lee2019mixed}
M.~Lee, N.~Norouzi, G.~Bruder, P.~J. Wisniewski, and G.~F. Welch.
\newblock Mixed reality tabletop gameplay: Social interaction with a virtual
  human capable of physical influence.
\newblock {\em IEEE Transactions on Visualization and Computer Graphics},
  27(8):3534--3545, 2019.

\bibitem{llobera2010proxemics}
J.~Llobera, B.~Spanlang, G.~Ruffini, and M.~Slater.
\newblock Proxemics with multiple dynamic characters in an immersive virtual
  environment.
\newblock {\em ACM Transactions on Applied Perception}, 8(1):1--12, 2010.

\bibitem{olckers2013psychological}
C.~Olckers.
\newblock Psychological ownership: Development of an instrument.
\newblock {\em SA Journal of Industrial Psychology}, 39(2):1--13, 2013.

\bibitem{scikit-learn}
F.~Pedregosa et~al.
\newblock Scikit-learn: Machine learning in {P}ython.
\newblock {\em Journal of Machine Learning Research}, 12:2825--2830, 2011.

\bibitem{pierce2001toward}
J.~L. Pierce, T.~Kostova, and K.~T. Dirks.
\newblock Toward a theory of psychological ownership in organizations.
\newblock {\em Academy of Management Review}, 26(2):298--310, 2001.

\bibitem{pierce2003state}
J.~L. Pierce, T.~Kostova, and K.~T. Dirks.
\newblock The state of psychological ownership: Integrating and extending a
  century of research.
\newblock {\em Review of General Psychology}, 7(1):84--107, 2003.

\bibitem{piumsomboon2018mini}
T.~Piumsomboon, G.~A. Lee, J.~D. Hart, B.~Ens, R.~W. Lindeman, B.~H. Thomas,
  and M.~Billinghurst.
\newblock Mini-me: An adaptive avatar for mixed reality remote collaboration.
\newblock In {\em Proceedings of the CHI Conference on Human Factors in
  Computing Systems}, pp. 1--13, 2018.

\bibitem{poretski2021owns}
L.~Poretski, O.~Arazy, J.~Lanir, and O.~Nov.
\newblock Who owns what? psychological ownership in shared augmented reality.
\newblock {\em International Journal of Human-Computer Studies}, 150:102611,
  2021.

\bibitem{poretski2019virtual}
L.~Poretski, O.~Arazy, J.~Lanir, S.~Shahar, and O.~Nov.
\newblock Virtual objects in the physical world: Relatedness and psychological
  ownership in augmented reality.
\newblock In {\em Proceedings of the CHI Conference on Human Factors in
  Computing Systems}, pp. 1--13, 2019.

\bibitem{libSyncPro}
\relax{Rogo Digital}.
\newblock {LipSync Pro}.
\newblock \url{https://rogodigital.gitbook.io/lipsync-pro/}.
\newblock Accessed: 2023-06-05.

\bibitem{PokemonGo}
\relax{The Pokemon Company}.
\newblock {Pok{\'e}mon Go}.
\newblock \url{https://pokemongolive.com}.
\newblock Accessed: 2023-02-22.

\bibitem{rudmin1987semantics}
F.~W. Rudmin and J.~W. Berry.
\newblock Semantics of ownership: A free-recall study of property.
\newblock {\em The Psychological Record}, 37(2):257--268, 1987.

\bibitem{sanz2015virtual}
F.~A. Sanz, A.-H. Olivier, G.~Bruder, J.~Pettr{\'e}, and A.~L{\'e}cuyer.
\newblock Virtual proxemics: Locomotion in the presence of obstacles in large
  immersive projection environments.
\newblock In {\em Proceedings of the IEEE Virtual Reality}, pp. 75--80, 2015.

\bibitem{van2004psychological}
L.~Van~Dyne and J.~L. Pierce.
\newblock Psychological ownership and feelings of possession: Three field
  studies predicting employee attitudes and organizational citizenship
  behavior.
\newblock {\em Journal of Organizational Behavior: The International Journal of
  Industrial, Occupational and Organizational Psychology and Behavior},
  25(4):439--459, 2004.

\end{thebibliography}
\end{document}